\documentclass[journal]{IEEEtran}
\usepackage{cite}

\ifCLASSINFOpdf
  \usepackage[pdftex]{graphicx}
  \graphicspath{{./figures/}}
  \DeclareGraphicsExtensions{.pdf,.jpeg,.png}
\else
  \usepackage[dvips]{graphicx}
  \graphicspath{{./figures/}}
  \DeclareGraphicsExtensions{.eps}
\fi
\usepackage{amsmath}
\def\etal{\emph{et al.~}}
\usepackage{graphicx}
\usepackage{epsfig}
\usepackage{epstopdf}
\usepackage{booktabs}       
\usepackage{amsmath}
\usepackage{amssymb}
\usepackage{multirow}
\usepackage{capt-of}
\usepackage{makecell}
\usepackage{fixltx2e}
\usepackage{capt-of}
\usepackage{siunitx}

\usepackage[pdftex, pdfauthor={Philipp Seeboeck}, pdftitle={Exploiting Epistemic Uncertainty of Anatomy Segmentation for Anomaly Detection in Retinal OCT}, pdfsubject={Biomarker Discovery in Retinal OCT Images}, pdfkeywords={weakly supervised learning, anomaly detection, biomarker discovery, optical coherence tomography, epistemic uncertainty}, pdfproducer={Latex with hyperref}, pdfcreator={pdflatex}]{hyperref}

\usepackage[normalem]{ulem}
\usepackage[dvipsnames]{xcolor}

\begin{document}
%
\title{Exploiting Epistemic Uncertainty of Anatomy~Segmentation for Anomaly~Detection in~Retinal~OCT}
%
%
%

\author{Philipp~Seeb{\"o}ck*,
	Jos{\'e} Ignacio Orlando,
	Thomas~Schlegl,
	Sebastian~M.~Waldstein,
	Hrvoje~Bogunovi{\'c},
	Sophie~Klimscha,
	Georg Langs*,
	and~Ursula~Schmidt-Erfurth
	\thanks{Copyright (c) 2019 IEEE. Personal use of this material is permitted. However, permission to use this material for any other purposes must be obtained from the IEEE by sending a request to pubs-permissions@ieee.org.}
	\thanks{Manuscript received January,10,2019}
	\thanks{The financial support by the Christian Doppler Research Association, the Austrian Federal Ministry for Digital and Economic Affairs and the National Foundation for Research, Technology and Development, by the Austrian Science Fund (FWF I2714-B31) is gratefully acknowledged. We thank the NVIDIA Corporation for GPU donation. J.I.O. is funded by WWTF AugUniWien/FA7464A0249 (Medical University of Vienna); VRG12-009 (University of Vienna).}
	\thanks{P.~Seeb{\"o}ck and G. Langs are with the Computational Imaging Research Lab, Department of Biomedical Imaging and Image-guided Therapy, Medical University Vienna, Austria  (email: \mbox{philipp.seeboeck@meduniwien.ac.at}, \mbox{georg.langs@meduniwien.ac.at})}
	\thanks{P.~Seeb{\"o}ck, J. I. Orlando, T.~Schlegl, S.M. Waldstein, H.~Bogunovic, S.~Klimscha, G. Langs and U. Schmidt-Erfurth are with the Christian Doppler Laboratory for Ophthalmic Image Analysis, Vienna Reading Center, Department of Ophthalmology and Optometry, Medical University Vienna, Austria. (email: \mbox{philipp.seeboeck@meduniwien.ac.at})}
	\thanks{* corresponding authors}}
	
%
%
%

\markboth{Accepted for publication in IEEE Transactions on Medical Imaging 2019}
{Seeb{\"o}ck \MakeLowercase{\textit{et al.}}: Exploiting Epistemic Uncertainty of Anatomy Segmentation for Anomaly Detection in Retinal OCT}%

%



\maketitle

\begin{abstract}
Diagnosis and treatment guidance are aided by detecting relevant biomarkers in medical images. 
Although supervised deep learning can perform accurate segmentation of pathological areas, it is limited by requiring \textit{a-priori} definitions of these regions, large-scale annotations, and a representative patient cohort in the training set. In contrast, anomaly detection is not limited to specific definitions of pathologies and allows for training on healthy samples without annotation. Anomalous regions can then serve as candidates for biomarker discovery. \\
Knowledge about normal anatomical structure brings implicit information for detecting anomalies. We propose to take advantage of this property using bayesian deep learning, based on the assumption that epistemic uncertainties will correlate with anatomical deviations from a normal training set. A Bayesian U-Net is trained on a well-defined healthy environment using weak labels of healthy anatomy produced by existing methods. At test time, we capture epistemic uncertainty estimates of our model using Monte Carlo dropout. 
A novel post-processing technique is then applied to exploit these estimates and transfer their layered appearance to smooth blob-shaped segmentations of the anomalies.
We experimentally validated this approach in retinal optical coherence tomography (OCT) images, using weak labels of retinal layers. Our method achieved a Dice index of 0.789 in an independent anomaly test set of age-related macular degeneration (AMD) cases. The resulting segmentations allowed very high accuracy for separating healthy and diseased cases with late wet AMD, dry geographic atrophy (GA), diabetic macular edema (DME) and retinal vein occlusion (RVO). Finally, we qualitatively observed that our approach can also detect other deviations in normal scans such as cut edge artifacts.
\end{abstract}

\begin{IEEEkeywords}
weakly supervised learning, anomaly detection, biomarker discovery, optical coherence tomography, epistemic uncertainty.
\end{IEEEkeywords}

%
\IEEEpeerreviewmaketitle

\section{Introduction}
\label{introduction}

Biomarker detection in medical imaging data plays a critical role in the context of disease diagnosis and treatment planning~\cite{nimse2016biomarker}.
However, performing this task manually is extremely expensive and time consuming. Moreover, as it requires experts in the field to know every possible visual appearance of the regions of interest, results may suffer from intra- and inter-grader variability~\cite{asman2011robust}.
Automated methods can partially address these issues by exploiting the potential of deep learning~\cite{litjens2017survey}. Supervised learning approaches are trained to detect well-known, pre-defined biomarker categories such as lesions or pathological changes in organs and tissues~\cite{esteva2017dermatologist,kooi2017large,rajpurkar2017chexnet,de2018clinically}. In retinal OCT imaging, supervised methods have been extensively used \cite{schmidt2018artificial}, e.g. for segmentation of fluid~\cite{schlegl2018fully,lee2018automated}, drusen~\cite{zadeh2017cnns}, hyperreflective material~\cite{schlegl2018fullyHRF} or photoreceptor disruptions~\cite{orlando2019u2}. However, these methods require large-scale annotated data sets, which can be costly or even unfeasible to obtain in some clinical scenarios. Moreover, their outputs are limited to the pre-defined set of marker categories, and are unable to discover novel biomarkers different from those used for training~\cite{seebock2018identify}.

\begin{figure}[t!]
	\centering
	\includegraphics[width=0.98\columnwidth]{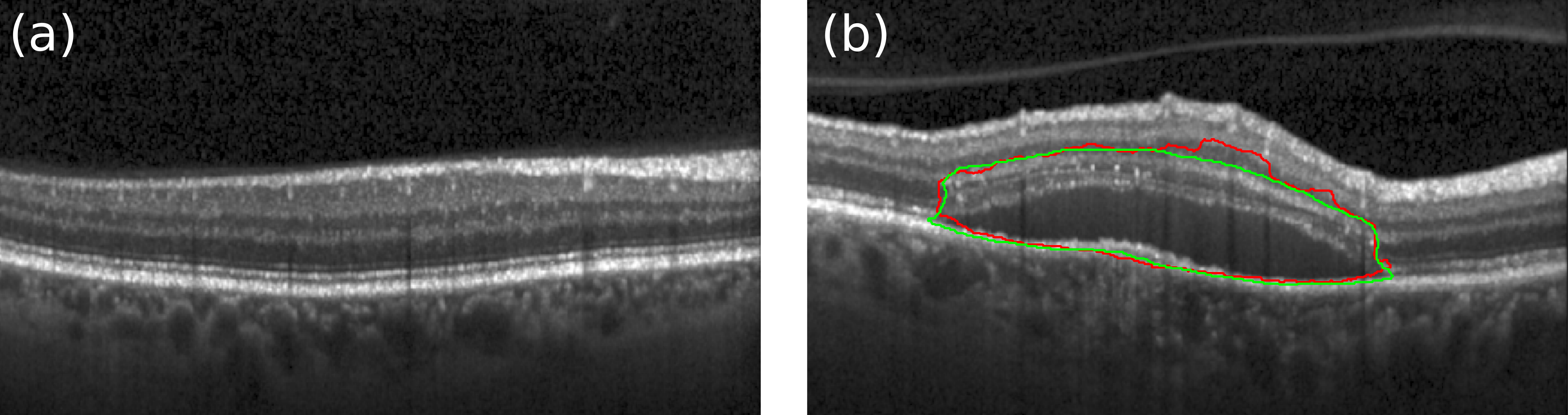}
	\caption{Anomaly detection in retinal OCT. (a) Healthy retina. (b) Diseased subject (manual annotation of anomaly in green, prediction of anomaly by our model in red).}
	\label{fig:results-intro}
\end{figure}

Anomaly detection methods offer an interesting alternative to supervised learning in this domain, as they are not limited in their application to a specific disease or marker category. Instead, these approaches leverage the knowledge extracted from healthy data during training, omitting the need of a representative patient cohort with an appropriate amount and variations of pathologies~\cite{pimentel2014review,editorial2018towards}. Capturing all possible disease related appearances or rare disease manifestations is costly or even unfeasible. In general, anomaly detection can be defined as a two-step process in which we first learn a model of normal appearance, and then we apply it to detect deviations from this normal data (anomalies) during test time~\cite{pimentel2014review,schlegl2017unsupervised,seebock2018identify}.
Therefore, instead of searching in the entire image space, these segmented anomalies can be explored by clinicians to identify features that might result in novel biomarkers, allowing a more efficient discovery process. Furthermore, identifying anomalous areas can be also helpful to efficiently screen for diseased cases in large patient cohorts.

Bayesian deep learning has emerged as an active field of research that aims to develop efficient tools for quantifying uncertainties~\cite{gal2015bayesian,kendall2015bayesian,kendall2017uncertainties}. In general, uncertainties can be classified into two main categories: \textit{aleatoric} and \textit{epistemic}. \emph{Aleatoric} uncertainty captures the vagueness inherent in the input, while \emph{epistemic} uncertainty refers to the model incertitude and can be reduced by incorporating additional data into the training process~\cite{kendall2017uncertainties}. Both aleatoric and epistemic uncertainty have previously been used for semantic segmentation \cite{kendall2015bayesian,kendall2017uncertainties}. In this context it has been shown that aleatoric uncertainty does not increase for examples different from those in the training set, while epistemic uncertainty does~\cite{kendall2017uncertainties}. Hence, the latter is more suitable for detecting changes (or anomalies) from the normal samples. Furthermore, disease classification methods based on deep learning were observed to be benefited by the usage of epistemic uncertainties~\cite{leibig2017leveraging}.

In this paper, we introduce a novel approach for anomaly detection, exploiting segmentation models of normal anatomy and their epistemic uncertainty while segmenting new images. Our method is based on the assumption that these uncertainties will correlate with deviations from a normal appearance.
We learn the regularities in the anatomy of healthy data, using weak labels. In this work we use the term "weak supervision" to indicate that we trained our model using labels automatically generated by a surrogate segmentation method instead of a human reader. We exploit this characteristic as traditional algorithms--even if they are not based on machine learning--are expected to perform accurately due to the well-defined properties of normal cases. Therefore, our approach does not involve manual labels at any stage. This setting allows to produce more training data and thereby to harvest more appearance variability.

We experimentally evaluate our approach in the context of anomaly detection in retinal optical coherence tomography (OCT) scans (Fig.~\ref{fig:results-intro}, Section~\ref{subsec:intro-oct}). We train a Bayesian U-Net~\cite{ronneberger2015u,joltikov2018disorganization} on a set of healthy images using weak labels of the retinal layers, provided by a standard graph-based method for layer segmentation~\cite{garvin2009automated}. At test time, we capture the epistemic uncertainty estimates from our network by means of Monte Carlo (MC) dropout~\cite{gal2015bayesian,kendall2015bayesian}. This output is postprocessed using a novel \emph{majority-ray-casting} technique in order to retrieve compact, blob-shaped smooth segmentations of the anomalies. On a separate test set of patients with age-related macular degeneration (AMD), our method achieves a Dice index of 0.789, outperforming previously published work by a large margin.
Furthermore, the performance of the proposed method is evaluated in a volume-level classification experiment, using only the amount of anomalous area as (discriminative) feature. By individually comparing healthy cases vs. diabetic macular edema (DME), retinal vein occlusion (RVO), dry geographic atrophy (GA) and late wet AMD, we observe that even this simple predictor allows to achieve almost perfect separation.

\subsection{Retinal OCT imaging}
\label{subsec:intro-oct}
OCT is a non-invasive volumetric imaging technique that provides high resolution images of the retina and is currently one of the most important diagnostic modalities in ophthalmology \cite{fujimoto2016oct}. A 3D OCT volume is composed of several 2D cross-sectional slices--or B-scans--, which are analyzed by physicians to determine treatments, diagnosis and other clinical decisions~\cite{fujimoto2016oct}.
Age-related macular degeneration (AMD) is one of the leading causes of blindness in the world~\cite{wong2014global}. Detectable AMD-related changes in OCTs are, among others, drusen, intra- and subretinal fluid, pigment epithelial detachment (PED) and photoreceptor loss~\cite{schmidt2018artificial}. Besides neovascular AMD, which is defined by the occurrence of fluid, geographic atrophy (GA) is the second form of late AMD, characterized by the death of retinal pigment epithelium (RPE) cells, photoreceptors and/or choriocapillaris. Other retinal diseases such as retinal vein occlusion (RVO)~\cite{jonas2010retinal} and diabetic macular edema (DME)~\cite{TAN2017143} are characterized by the occurrence of intraretinal/subretinal fluid. Presence or changes in some of these features have been shown to correlate with visual function or disease progression~\cite{vogl2017analyzing}. Predictive capability however remains to be limited and underlying pathogenetic mechanisms are not yet fully understood~\cite{SchmidtErfurth20161}, meaning that there might be other unknown structures or patterns that are still needed to be discovered.

We propose to apply our uncertainty based approach to automatically segment anomalies in retinal OCT scans. In this domain, \emph{normal} is defined as the absence of pathological changes beyond age-related alterations. According to the Beckman Initiative Classification~\cite{ferris2013clinical}, we allowed drusen below 63 $\mu m$ in size as only visible alteration, as they normally do not result in visual impairment.
A set of healthy retinas and corresponding weak labels obtained using~\cite{garvin2009automated} are used to train a Bayesian deep learning model for segmenting the retinal layers. Pixel-wise epistemic uncertainty estimates are applied at test time to identify anomalous regions in new given samples. While pathologies such as subretinal fluid are known to alter the appearance of the retina, some other are strictly related with the layers (e.g. the disorganization of the retinal inner layers, or DRIL)~\cite{joltikov2018disorganization}. Therefore, using retinal layer information is an appropriate way of incorporating anatomical knowledge into the model.
At the same time, no labels of the target class (i.e. anomalies) are needed for training.

\begin{figure*}[t!]
	\centering
	\includegraphics[width=0.98\textwidth]{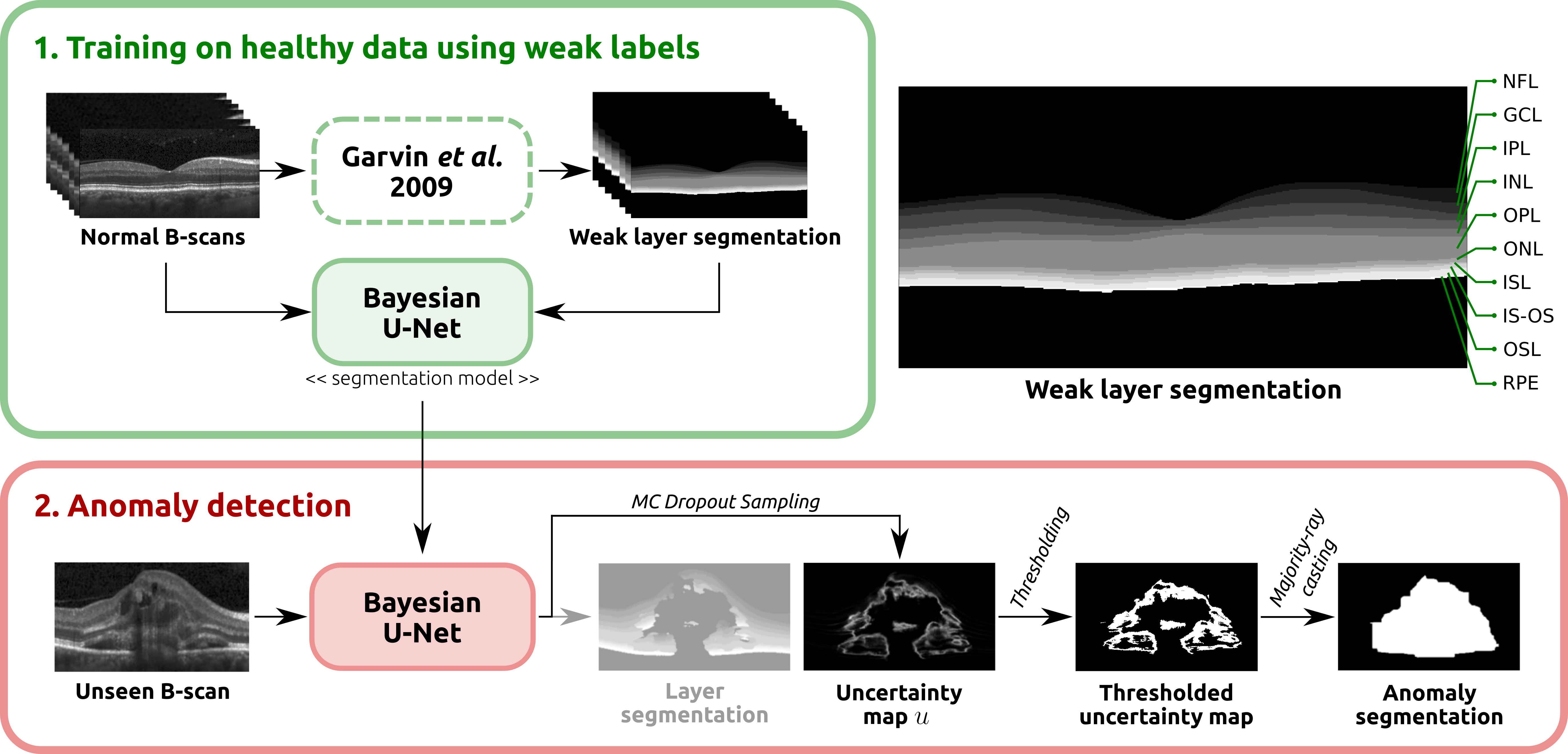}
	\caption{Overview of the proposed method. A Bayesian U-Net is trained on \emph{normal} B-scans, using weak labels of the retinal layers generated using the Garvin \textit{et al.}~\cite{garvin2009automated} segmentation method. The retinal layers are indicated on the right hand side. Given an unseen B-scan, MC dropout sampling is used to retrieve epistemic uncertainty maps, which are subsequently post-processed using \emph{majority-ray-casting} to obtain the final anomaly segmentations.} 
	\label{fig:method_overview}
\end{figure*}

\subsection{Related Work}
\label{introduction:relatedwork}

Biomarker discovery and analysis have benefited by the incorporation of deep learning~\cite{poplin2018prediction}. Non data-driven approaches require hand-crafting techniques to capture a specific biomarker, and then assess its statistical power, e.g. by means of linear discriminant analysis~\cite{orlando2018towards}. Alternatively, supervised deep learning avoids biases due to manual design of features by learning them from data. These techniques have been extensively used to identify pre-defined pathological markers such as disease lesions~\cite{esteva2017dermatologist,nair2018exploring,schmidt2018artificial,de2018clinically}. Their main drawback is that they require a training set with manual annotations of the region of interest. Thus, the markers have to be known in advance--restricting the possibility of using these models for biomarker discovery--and the training set has to include enough representative examples of the marker appearances. Alternatively, some authors propose to predict pre-defined clinical parameters using image-based regression techniques~\cite{poplin2018prediction,kermany2018identifying}. These methods assume that the networks will learn to capture features in the images that are correlated with relevant target values. Appropriate visualization techniques are needed to understand the properties of the model and to identify the features taken into account for prediction~\cite{poplin2018prediction,kermany2018identifying}. The regression target needs to be pre-defined, and it can be either a functional parameter~\cite{poplin2018prediction} or a diagnosis~\cite{kermany2018identifying}. Furthermore, a representative sample of diseased subjects has to be included in the training set if the target parameters are related to a specific condition. Moreover, due to the complexity of the prediction task, a larger number of training samples is required compared to supervised segmentation approaches.

Anomaly detection, on the contrary, identifies pathological areas that are implicitly defined by healthy data: normal appearance is first learned from this data, and anomalies are obtained in new data by detecting the difference to this representation. This overcomes the need of a sufficiently representative cohort of diseased patients, to select features with stable predictive value for a given target. Instead, first anomalies are detected based on a model trained on large-scale healthy data, and highlighted in the images as blob-shaped segmentations. In a second step, these candidates--typically only a fraction of the overall data--can be mined more efficiently for discovering new biomarkers and/or predictors.
These techniques can be applied as a first step in discovering novel risk factors of diseases, extending the vocabulary of known biomarkers, and therefore our knowledge about the underlying pathogenesis of diseases \cite{schlegl2017unsupervised,pawlowski2018unsupervised,seebock2018identify}.

Multiple techniques have been proposed in the past for automated anomaly detection in OCT images~\cite{dufour2012pathology,schlegl2017unsupervised,sidibe2017anomaly,seebock2018identify}.
Shape models were used in~\cite{dufour2012pathology} to perform drusen detection. In~\cite{sidibe2017anomaly}, the appearance of normal OCT B-scans was modeled with a Gaussian Mixture Model (GMM), recognizing anomalous B-Scans as outliers. Entire OCT volumes were classified as normal or anomalous, based on the number of outliers. 
Deep unsupervised anomaly detection has been recently presented in \cite{schlegl2017unsupervised,seebock2018identify}, both relying on a representation learned at patch-level. Schlegl~\etal~\cite{schlegl2017unsupervised} used a Generative Adversarial Network (GAN) to learn a manifold of normal anatomical variability, and anomalies were detected as deviations from it. A multi-scale autoencoder approach combined with a one-class support vector machine (SVM) was presented in~\cite{seebock2018identify} to segment anomalies and to identify disease clusters subsequently.
None of these anomaly detection approaches incorporate the use of uncertainty.

To the best of our knowledge, uncertainties were not used for anomaly detection before. In particular, Nair~\etal~\cite{nair2018exploring} used Bayesian supervised learning to segment multiple sclerosis lesions in MRI. Sedai~\etal~\cite{sedai2018joint} applied a similar method for layer segmentation in healthy OCT scans. In both works, aleatoric uncertainty was used for training. In~\cite{nair2018exploring}, epistemic uncertainty was applied to refine the segmentations, while in~\cite{sedai2018joint} the epistemic uncertainty was provided as qualitative feedback to users. Monte Carlo sampling with dropout was used in~\cite{pawlowski2018unsupervised} to average multiple outputs from an autoencoder trained in healthy data. Anomalies were detected as differences between the input and the reconstructed output. In this paper we aim for a different task compared to these previous approaches: we use the epistemic uncertainty of a model trained on healthy subjects to discover anomalies in new data.

\subsection{Contributions}

We propose a novel approach for anomaly detection based on epistemic uncertainty estimates from a Bayesian U-Net, trained for segmenting the anatomy of healthy subjects. To the best of our knowledge, this is the first method to pose the segmentation of anomalies in this way. In addition, our model is trained on weak labels instead of manual annotations, which allows to increase the training data without major efforts. We evaluate our model in the context of anomaly detection in retinal OCT scans. We introduce a heuristical post-processing technique, namely majority-ray-casting, to ensure compact-shape consistency in the final anomaly segmentations. Our approach is able to obtain a clear performance improvement compared to previous state-of-the-art in anomaly detection in OCTs~\cite{seebock2018identify}. This manifests in a Dice of $0.789$ on the anomaly test set regarding the pixel-wise segmentation task, while achieving almost perfect volume-level separation of healthy and diseased volumes with late wet AMD, dry GA, DME and RVO, solely based on the area of detected anomalies. Finally, we also qualitatively observed high uncertainty estimates in regions with other deviations such as imaging artifacts in normal subjects.

\section{Methods}
\label{sec:method}

An overview of the proposed approach is illustrated in Fig.~\ref{fig:method_overview}. First, we train a Bayesian U-Net model on \emph{normal} cases to segment retinal layers, using weak labels automatically generated with a graph-based segmentation approach. Secondly, this model is applied together with Monte Carlo dropout~\cite{kendall2015bayesian,gal2015bayesian} to retrieve pixel-level epistemic uncertainty estimates. Finally, we introduce a simple post-processing step, \emph{majority-ray-casting}, to transform the uncertainty maps into compact segmentations of anomalies. This technique closes the gap between the shape of layers and anomalies based on the assumption that anomalies in OCT are compact and not layered.

Section~\ref{subsec:method_healthy} describes the general idea of training a segmentation model from a healthy population using weak labels. Section~\ref{subsec:method_uncertainty} focuses on the application of the epistemic uncertainty estimates of this model for anomaly detection. The domain-specific pipeline for applying the anomaly detection approach in retinal OCT scans is presented in Section~\ref{subsec:application-oct}

\begin{figure*}[t]
	\centering
	\includegraphics[width=0.9\textwidth]{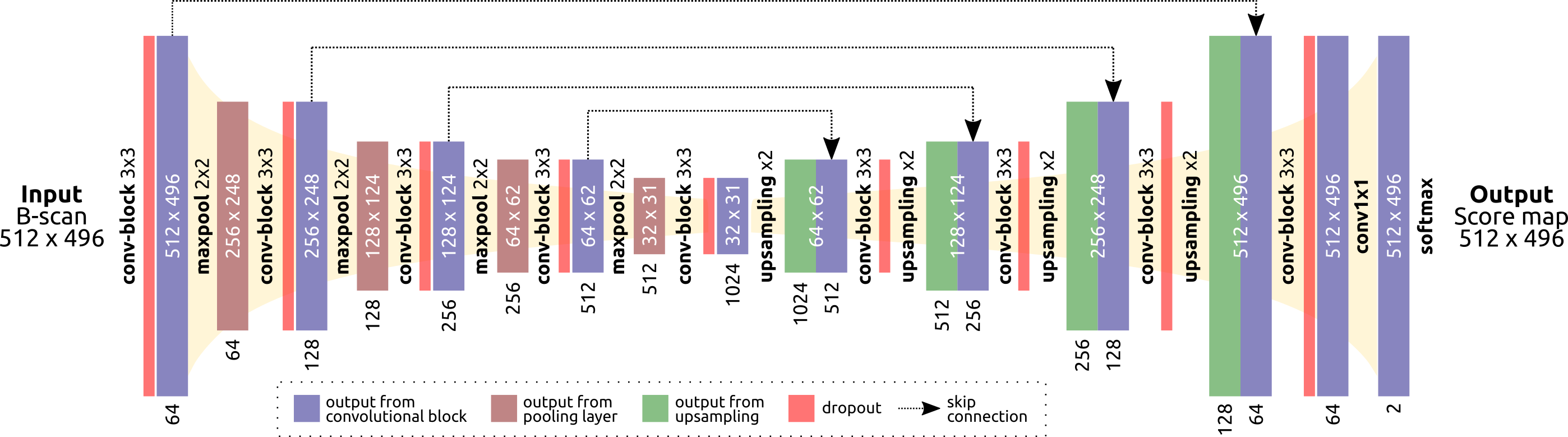}
	\caption{Overview of the network architecture. Each convolutional block has the following structure: 3-by-3 convolution + batch-normalization + ReLU + 3-by-3 convolution + batch-normalization + ReLU. All convolutional layers use a stride of 1 and zero padding. A combination of nearest neighbor upsampling and a convolutional layer is applied instead of transposed convolutions.}
	\label{fig:network_architecture}
\end{figure*}

\subsection{Training on Healthy Population}
\label{subsec:method_healthy}

Let $ X \in \mathbb{R}^{a \times b} $ be a set of \emph{normal} images with $a \times b$ pixels size, and $ Y \in \mathcal{Y}^{a \times b} $ the set of corresponding weak, target label maps, with $\mathcal{Y} = \{1, ..., K\}$ the set of all possible classes. A segmentation model aims at finding the function $f_W: X \rightarrow Y$ by optimizing its set of weights $W$.
In this study, we model $f_W$ using a multiclass U-Net~\cite{ronneberger2015u}. This widely used segmentation architecture is composed of an encoding and a decoding part with skip-connections: the encoder contracts the resolution of the input image and captures the context and relevant features on it, while the decoder performs up-sampling operations to enable precise localization of the target class and restores the input resolution. The skip-connections, on the other hand, allow to better reconstruct the final segmentation by transferring feature maps from one encoding block to its counterpart in the decoder. Our instance of the U-Net (Fig.~\ref{fig:network_architecture}) comprises five levels of depth, with $64$, $128$, $256$, $512$ and $1024$ output channels each. Dropout is applied after each convolutional block, which consists of two $3 \times 3$ convolutions, each followed by batch-normalization~\cite{ioffe2015batch} and a rectified linear unit (ReLU). $2 \times 2$ max-pooling and nearest-neighbor interpolation are used for downsampling and upsampling, respectively. The network is trained with the cross entropy loss objective function.

\subsection{Exploiting Epistemic Uncertainty for Anomaly Detection}
\label{subsec:method_uncertainty}

Epistemic uncertainty was observed to increase when estimated on image samples whose appearance differ significantly from those on the training data~\cite{kendall2017uncertainties}. We propose to exploit this characteristic to identify and segment anomalies in unseen scans. 

Formally, Bayesian deep learning aims to find the posterior distribution over the weights of the network $p(W|X,Y)$, in order to derive epistemic uncertainty. In general, retrieving the actual true underlying distribution is computationally intractable, so it needs to be approximated. Gal~\etal~\cite{gal2015bayesian} proposed to approximate the posterior with the variational distribution $q(W)$, i.e. by using dropout also at test time to retrieve MC samples. This is theoretically equivalent to modelling $q$ as a Bernoulli distribution with probability $p$ equal to the dropout rate.
It has been shown in~\cite{gal2015bayesian} that the Kullback-Leibler divergence between the approximate and posterior distribution:
\begin{equation}
KL(q(W)||p(W|X,Y))
\end{equation}
is minimized by optimizing the cross-entropy loss during training. Hence, training the network with gradient descent and dropout not only prevents over-fitting but also encourages the network to learn a weight distribution that properly explains the training data.

At test time, given an \emph{unseen} image $x$ (e.g. a B-scan), the pixel-wise epistemic uncertainty is estimated as follows. First, $n$ predictions $\mathbf{y}^{(i)}$, $i \in 1,\dots,n$ are retrieved by applying the model $f_{W \sim q(W)}$ on $x$. The pixel-wise variance $\mathbf{\sigma}^2$ is then computed for each class $k \in \mathcal{Y}$ by:
\begin{equation}
    \sigma_{k}^2(\mathbf{p}) = \frac{1}{n}  \sum^{n}_{i} {\left( y^{(i)}_{k}(\mathbf{p}) - \mu_{k}(\mathbf{p}) \right)}^2
\end{equation}
where $\mathbf{p}$ is a pixel coordinate and $\mu_k$ is the average of the $n$ predictions for the $k$-th class. The final uncertainty map $u$ is obtained by averaging all $\sigma_{k}^2$ estimates over the $K$ class-specific variances in a pixel-wise manner:
\begin{equation}
    {u(\mathbf{p}) = \frac{1}{K}  \sum^{K}_{k} \sigma_{k}^2(\mathbf{p})}.
\end{equation}

\subsection{Application of anomaly detection in retinal OCT scans}
\label{subsec:application-oct}

We apply the uncertainty-based anomaly detection approach to retinal OCT scans. The training set consists of pairs $(X,Y)$ composed of a healthy OCT B-scan $X$ and its associated weak labelling map $Y$ of the retinal layers. $Y$ is pre-computed using the graph-based surface segmentation algorithm described in~\cite{garvin2009automated}. Such a method has proven to be effective in normal subjects and is widely applied in ophthalmological studies~\cite{vogl2017predicting,bogunovic2017machine}. The set $\mathcal{Y}$ of labels comprises $K=11$ classes corresponding to background and $10$ retinal layers (Figure~\ref{fig:method_overview}): nerve fiber layer (NFL); ganglion cell layer (GCL); inner plexiform layer (IPL); inner nuclear layer (INL); outer plexiform layer (OPL); outer nuclear layer (ONL); inner segment layer (ISL); inner segment - outer segment (IS-OS) junction; outer segment layer (OSL) and the retinal pigmented epithelium (RPE). 

We use these weak labels to train the Bayesian multiclass U-Net described in Section~\ref{subsec:method_healthy}. The neural network provides both a segmentation map and an uncertainty estimate. We only use the latter at test time, as our purpose is not to accurately identify the retinal layers but to segment retinal abnormalities.

\begin{figure}[t!]
	\centering
	\includegraphics[width=0.99\columnwidth]{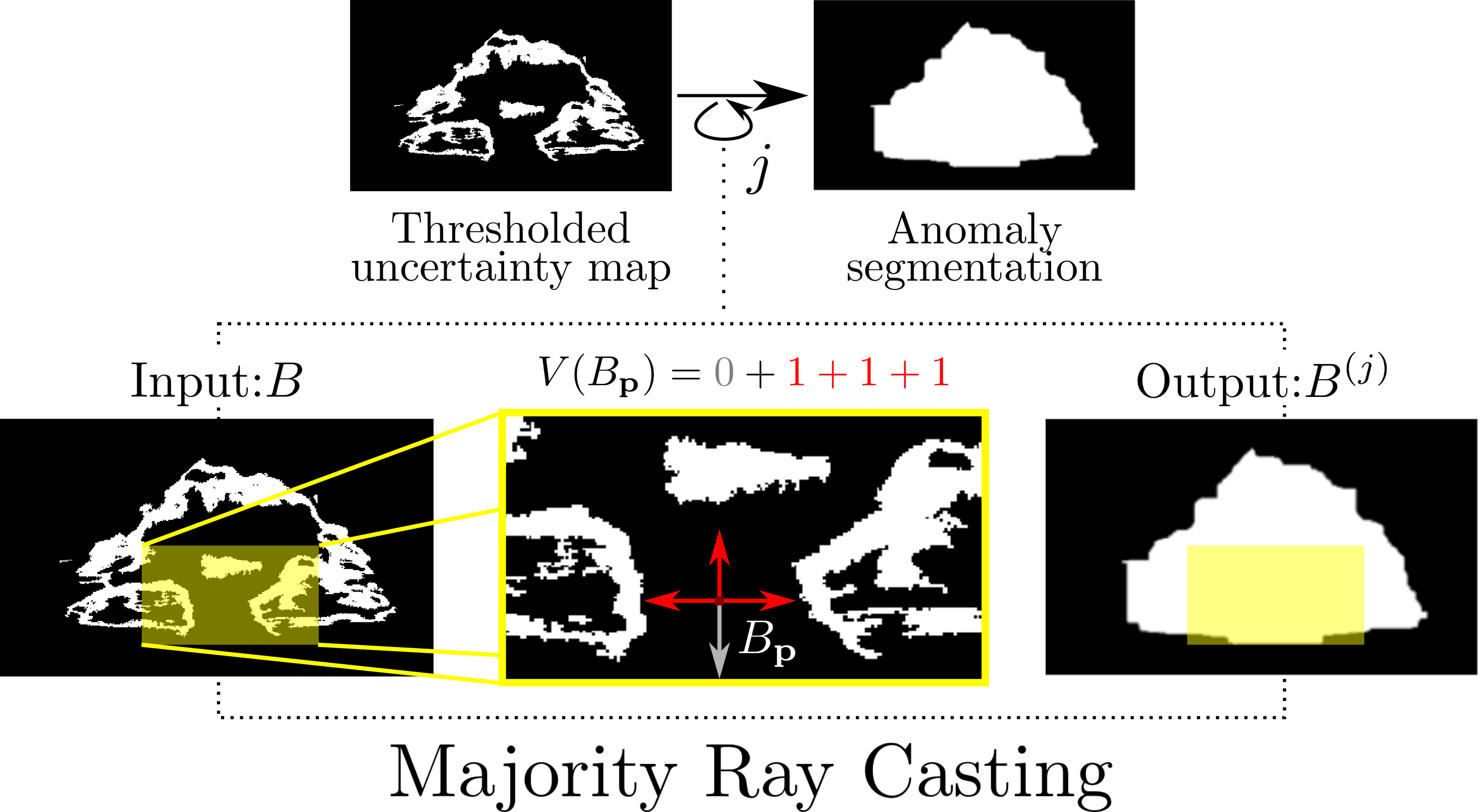}
	\caption{Majority-ray-casting post-processing technique. Red arrows indicate ray hits with a binary (white, =1) region, while the gray arrow indicates a non-hit.}
	\label{fig:majority-ray-casting}
\end{figure}

A first estimate of the anomalous areas is obtained by thresholding $u$ with a threshold $t$. To eliminate spurious predictions, every connected component with an area smaller than $s$ pixels is removed, resulting in a binary map $B$. The most straightforward way to highlight anomalies in an input B-scan is by providing compact, blob-shaped smooth segmentations surrounding the abnormal areas. As can be seen Fig.~\ref{fig:majority-ray-casting}, $B$ is not smooth enough to fit that shape.

We introduce a simple but effective technique, \textit{majority-ray-casting}, that iteratively postprocesses the binary map $B$ and results in a more shape consistent anomaly segmentation. This approach assumes that the retina is approximately horizontally orientated in the B-scan, which is usually the case. A schematic representation of the method is provided in Fig.~\ref{fig:majority-ray-casting}.
On an iteration $j$, in a first step four rays are sent to each of the cardinal coordinates (left, right, top and bottom) from every pixel $\mathbf{p}$ that satisfies $B_{\mathbf{p}}=0$. In other words, each black pixel in Fig.~\ref{fig:majority-ray-casting} is used once as reference point to cast the four rays. Each ray that ''hits'' a pixel with value $1$ before reaching the border of $B$ increases a pixel-wise ray-casting vote $V(B_{\mathbf{p}})$ by $1$. Hence, the maximum voting value of $V(B_{\mathbf{p}})$ for each pixel $\mathbf{p}$ can be $4$.
In a second step, all pixels with votes greater than or equal to a hyper-parameter $v^{(j)}$ are then set to $1$, resulting in a new binary map $B^{(j)}$. Formally, this can be written as:
\begin{equation}
    B^{(j)}_{\mathbf{p}}=
	\begin{cases}
	1 & \text{if}\ B_{\mathbf{p}} = 1 \\
	1 & \text{if}\ V(B_{\mathbf{p}}) \geq v^{(j)} \\
	0 & \text{if}\ V(B_{\mathbf{p}}) < v^{(j)} .
	\end{cases}
\end{equation}
Notice that this process can be iteratively repeated using $B^{(j)}$ as an input to the next iteration, and a different value of $v^{(j)}$ can be used at each iteration. 
Finally, morphological closing and opening operations with a radius of $m_c$ and $m_o$, respectively, were applied to remove artifacts.

\section{Experimental setup}
\label{sec:experimental_setup}

We empirically evaluated our method in our application scenario. In particular, we studied: (1) if our method can accurately identify anomalous regions in retinal OCT data, (2) the contribution of each of the individual components of our proposed approach in the final results, (3) the lesion-wise detection performance of the method, and (4) the volume-wise classification accuracy of the algorithm, based on the average number of anomalous pixels per B-scan for each volume.

\paragraph{Data}
We used six data sets of macula centered Spectralis (Heidelberg Engineering, GER) OCT scans, with $512\times496\times49$ voxels per volume, covering approximately $6 \text{mm} \times 2 \text{mm} \times 6 \text{mm}$ of the retina.
The first two datasets \emph{normal} and \emph{normal evaluation} comprise 226 and 33 healthy volumes, respectively, which were selected from 482 / 209 contralateral eye scans of patients with Retinal Vein Occlusion (RVO) / AMD in the other eye. According to the definition of \emph{healthy} provided in Section~\ref{introduction}, volumes with pathological changes beyond age-related alterations were excluded. 
The \emph{normal} data set was randomly split on a patient basis into 198 training and 28 validation cases to train our segmentation model. The \emph{normal evaluation} set, on the other hand, was only used for evaluation purposes. 
The third dataset \emph{late wet AMD} comprised 31 OCT volumes ($5$ validation, $26$ test) with active neovascular AMD. A retina specialist manually annotated all the areas containing pathologic features, resulting in pixel-wise annotations of anomalous regions.
All these datasets have been already used for training and evalution in \cite{seebock2018identify}, using the same configuration. This allows a direct comparison with such an approach.\\
Four volume-wise disease classification experiments were performed by comparing the anomalous areas in healthy subjects vs diseased. The \textit{late wet AMD test set}, 30 DME, 25 RVO and 34 dry GA volumes were used separately for this purpose.

\paragraph{Training details}
Intensity values of each individual B-scan were rescaled between 0 and 1 before being fed to the network.
We used Kaiming initialization~\cite{he2015delving}, Adam optimization~\cite{kingma2014adam}, and a learning rate of $1 \times 10^{-4}$ 
(multiplied by $0.2$ every $5$ epochs). During training, random data augmentations were applied, including horizontal flipping, rotation up to \ang{10}, horizontal / vertical translations up to 5\% / 20\% of the B-scan size and scaling up to 2\%. The network was trained for $25$ epochs on the \emph{normal training set}, and the model with the best average Dice for layer segmentation on the \emph{normal validation set} was selected for evaluation.
We trained the model with different dropout rates $p=\{0.1, 0.2, 0.3, 0.4, 0.5\}$, and the model with the highest Dice for anomaly segmentation on the \emph{late wet AMD validation set} was selected for performance evaluation on the \emph{late wet AMD test set}.

\paragraph{Anomaly detection details}
At inference time, $50$ MC samples with dropout were retrieved per B-scan. For post-processing, we used $s=10$, $m_c=4$, and $m_o=2$, where these parameters were selected empirically by qualitatively analyzing the results in a few B-scans from the \emph{late wet AMD validation set}. Two iterations of the \emph{majority-ray-casting} algorithm were performed, using $v^{(1)}=3$ and $v^{(2)}=4$, and $20$ different thresholds $t=\{0.01, 0.02, ... 0.19, 0.20\}$ were evaluated on the \emph{lateAMD validation set}. The best threshold according to the average validation Dice was selected for performance evaluation on the \emph{lateAMD test set}. This calibration ensured to retrieve compact annotations consistent with the desired blob-shape appearance.

\subsection{Segmentation accuracy}
The segmentation accuracy was evaluated using precision, recall and Dice, which are standard metrics for binary segmentation tasks. Notice that performing a ROC curve based evaluation is unfeasible in our case as our method does not produce pixel-level likelihood predictions of anomalies, but binary labels.

To assess the contribution of each individual component of our proposed approach in the final results, we performed a series of ablation experiments. It is worth mentioning that the test set was not used for designing the method: all our design decisions were based on the validation set performance. These ablation studies are performed on the test set only to illustrate how changing our model can affect the results. For the sake of brevity, from now on we will refer to the full method described in Section~\ref{sec:method} as \emph{WeakAnD} (from \textit{\textbf{Weak} \textbf{An}omaly \textbf{D}etection}).
\begin{itemize}
	\item \emph{Binary layer-segmentation}: While the proposed \emph{WeakAnD} is trained with 11 layer classes, we trained a second network, namely \emph{WeakAnD(binary)}, for the binary segmentation task ''retina/background''. This experiment allows to assess the influence of annotation details in the anomaly detection performance.
	
	\item \emph{Remove majority-ray-casting}: To show the necessity of the majority-ray-casting approach, we compared against a simple post-processing only thresholding the uncertainty maps $u$ (\emph{WeakAnD (thresholding)}). We also replaced the majority-ray-casting step with a straightforward convex hull step (\emph{WeakAnD (convex-hull)}).
	
	\item \emph{Remove morphological operations}: The final morphological closing and opening operations were removed in this ablation experiment (\emph{WeakAnD (w/o closing/opening)}). 

	\item \emph{Layer flattening}: As an additional pre-processing step for the \emph{lateAMD} dataset, the retina was flattened using the bottom layer (Bruch's Membrane - BM), projecting it onto a horizontal plane, following the pre-processing approach in \cite{seebock2018identify}. Our hypothesis is that flattening the retina helps to meet the assumption of majority-ray-casting, i.e. horizontal orientation of the retina.
\end{itemize}

\subsection{Lesion-wise Detection}
We are interested in evaluating the detection performance of the proposed approach on a lesion-wise basis. To this end, we define each connected anomaly within a B-scan as a single lesion (e.g., Fig.~\ref{fig:ablationStudies-qualitative}(b) presents two lesions). A Dice index is computed for each individual lesion to quantify its overlap with its corresponding manual annotation. A thresholding according to a reference value $d$ is then performed, where the amount of true positives is counted as the number of lesions with a Dice index higher than $d$. These values are used to compute \textit{lesion-detection Recall (LD-Re$_d$)} and \textit{lesion-detection Precision (LD-Pr$_d$)}:
\begin{equation}
    \text{LD-Re}_d = \frac{\text{TP}_d}{\text{TP}_d + \text{FN}_d}
\end{equation}
\begin{equation}
    \text{LD-Pr}_d = \frac{\text{TP}_d}{\text{TP}_d + \text{FP}_d}
\end{equation}
where $\text{TP}_d$, $\text{FN}_d$ and $\text{FP}_d$ are the number of true positive, false negative and false positive lesions for a given $d$. By computing these metrics for each possible $d \in [0,1]$, we can then plot both \emph{LD-Re} and \emph{LD-Pr} curves. These plots allow to assess the stability of the Dice values with respect to the lesion detection performance. Notice that this cannot be used to select an operating point as it is defined over all possible dice values and not on lesion probabilities.

\subsection{Volume-wise Disease Detection}
We conducted four additional experiments to evaluate if the proposed method can be used to discriminate diseased versus healthy patients and to further assess the behavior of our approach on healthy cases. Without additional training, the average anomalous area per B-scan for each volume was directly used as a discriminative feature to separate between healthy and diseased cases. The following setups were used: \textit{normal evaluation} vs \textit{late wet AMD test set}, \textit{normal evaluation} vs GA, \textit{normal evaluation} vs RVO and \textit{normal evaluation} vs DME.

\section{Results}
\label{sec:results}

\begin{table}[t!]
	\centering
	\captionof{table}{Quantitative results on the late wet AMD validation set with varying dropout parameters.}
	\label{table:anomaly-detection-validation}
	\centering
	\resizebox{0.2\textwidth}{!}{
		\begin{tabular}{ll}
			\toprule
			Dropout &  Dice \\
			\midrule
			0.1     & 0.783 (0.05)  \\
			0.2     & 0.778 (0.02)  \\
			0.3     & 0.796 (0.03)  \\
			0.4     & 0.798 (0.03)  \\
			0.5     & 0.783 (0.03)  \\
			\bottomrule
		\end{tabular}
	}
\end{table}

\begin{table}[t!]
	\centering
	\captionof{table}{Quantitative results of anomaly detection on the late wet AMD test set.}
	\label{table:anomaly-detection}
	\centering
	\resizebox{0.48\textwidth}{!}{
		\begin{tabular}{llll}
			\toprule
			Method & Precision & Recall & Dice \\
			\midrule
			DDAE\textsubscript{ent} \cite{seebock2018identify}  & 0.47 (0.12) & 0.63 (0.06) & 0.53 (0.09) \\
			Entropy of Soft Predictions                         & 0.600 (0.08) & 0.622 (0.09) & 0.606 (0.07) \\
			\textbf{WeakAnD}                      				& \textbf{0.739 (0.06)} & \textbf{0.808 (0.07)} & \textbf{0.768 (0.03)} \\
			\textbf{WeakAnD} (with layer-flattening)             & \textbf{0.748 (0.06)} & \textbf{0.844 (0.07)} & \textbf{0.789 (0.03)} \\
			\bottomrule
		\end{tabular}
	}
\end{table}

\begin{figure*}[t]
	\centering
	\includegraphics[width=1.0\textwidth]{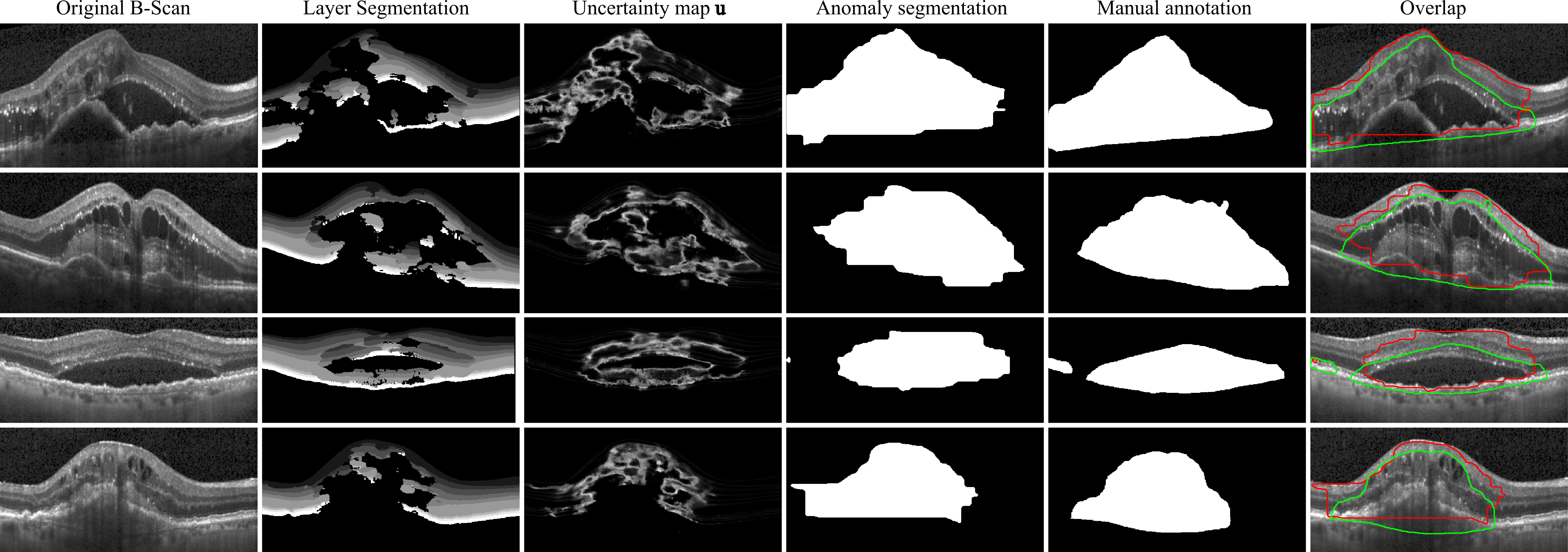}
	\caption{Qualitative results of the proposed method, on the late wet AMD test set. Central B-scans of volumes in which the proposed method performed best/worst in terms of the Dice index are shown. The corresponding Dice values are 0.82, 0.81, 0.72 and 0.72, from top to bottom. The last column indicates the overlap between the manual annotations of anomaly in green and the prediction of anomaly by our model in red.}
	\label{fig:qualitative_results}
\end{figure*}

\begin{figure}[t]
	\centering
	\includegraphics[width=0.3\textwidth]{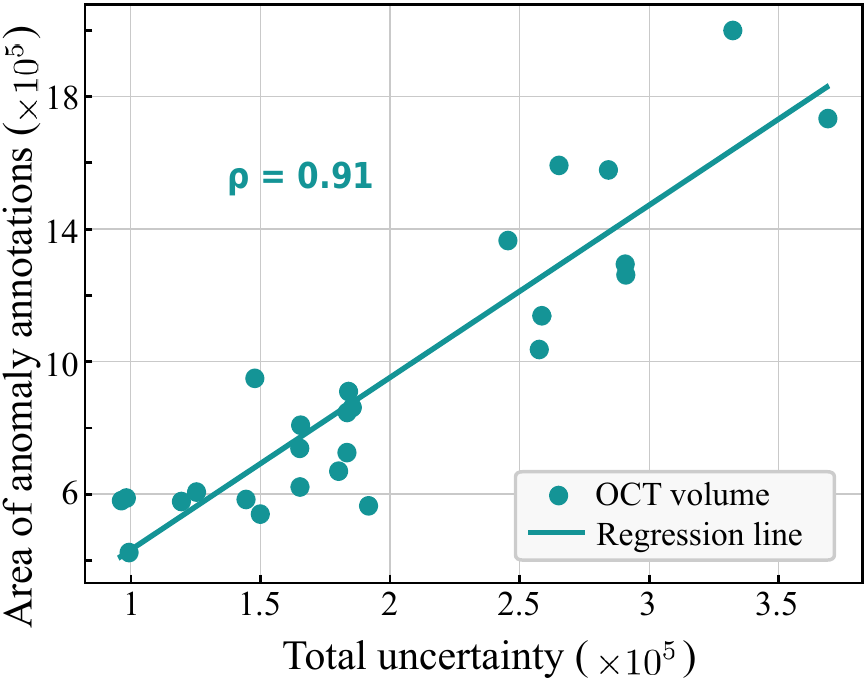}
	\caption{Correlation between the total amount of uncertainty and the area of anomaly annotations. Each point in the plot corresponds to one OCT volume in the \emph{late wet AMD test set}. The least-squared-fit line as well as the Pearson correlation coefficient $\rho$ are provided.}
	\label{fig:correlation-uncertainty-anomaly}
\end{figure}

Quantitative results for anomaly detection are provided in Table~\ref{table:anomaly-detection}. Two baselines are included: the state-of-the-art method described in~\cite{seebock2018identify} and an additional approach based on replacing our epistemic uncertainty estimates by the entropy of the soft predictions of the layers. Notice that majority-ray-casting was also applied to the entropy-based baseline to ensure a fair comparison. It can be seen that the proposed approach outperformed the two baselines by a large margin. When layer-flattening is applied to pre-process the OCT volumes as in \cite{seebock2018identify}, an improvement in performance is also observed, with a statistical significant increment in the Dice values from $0.768$ to $0.789$ (paired Wilcoxon signed-rank test, $p=0.00007$).
The final \emph{WeakAnD} model used a threshold of $t=0.10$ and a dropout rate of $p=0.4$. However, we experimentally observed that the performance on the validation set was not too sensitive to the dropout parameter (Table~\ref{table:anomaly-detection-validation}.

Qualitative anomaly segmentation results obtained in the \textit{late AMD test set} are shown in Fig.~\ref{fig:qualitative_results}. The central B-scans, corresponding to the volumes in which our method performed best/worst in terms of Dice, are shown in the top/bottom two rows. An additional example of a non central B-scan is depicted in Fig.~\ref{fig:results-intro}. Further qualitative results in DME, RVO and GA cases are depicted in Fig.~\ref{fig:healthyVSdiseased_qualitative} and in the supplementary material.

A scatter plot comparing the total area (in pixels) of anomalies (as manually annotated by the expert) and the level of uncertainty of the segmentation model is depicted in Fig.~\ref{fig:correlation-uncertainty-anomaly}. Each point corresponds to an individual OCT volume in the \emph{late wet AMD test set}. The linear regression line for the corresponding values is also included in the plot. The correlation between variables, as measured using the Pearson correlation coefficient, is $\rho = 0.91$.

\begin{figure}[t]
	\centering
	\includegraphics[width=0.48\textwidth]{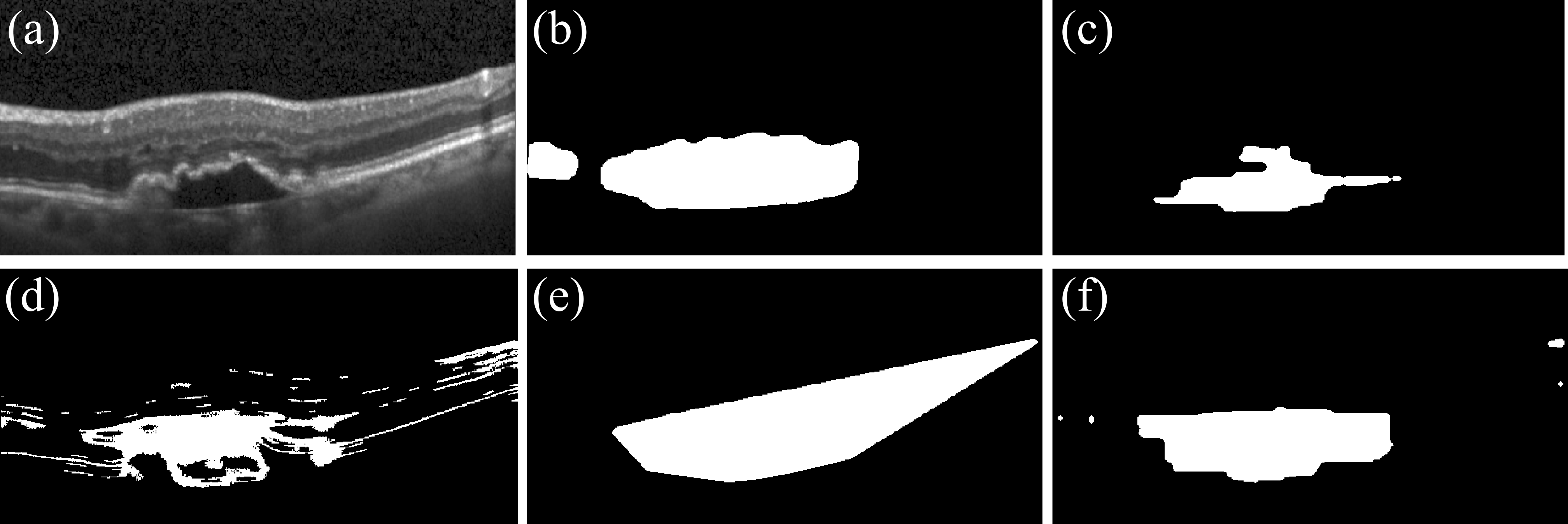}
	\caption{Qualitative results of the ablation studies, showing anomaly segmentation results on an exemplary sample. (a) Original B-scan, (b) Manual annotation, Segmentation results of (c) WeakAnD (binary), (d) WeakAnD (thresholding), (e) WeakAnD (convex-hull) and (f) WeakAnD.}
	\label{fig:ablationStudies-qualitative}
\end{figure}

\paragraph{Segmentation Accuracy}
Table~\ref{table:anomaly-detection-ablation} provides quantitative results of the conducted ablation studies, while qualitative results are shown in Fig.~\ref{fig:ablationStudies-qualitative}. It can be observed that all the ablations resulted in a performance loss, with different quantitative and qualitative effects. In particular, the importance of using a fine-grained layer segmentation is highlighted by the drop in the observed evaluation metrics when using a binary segmentation.

\begin{table}[t!]
	\centering
	\captionof{table}{Quantitative results of the ablation studies, as evaluated on the late wet AMD test set.}
	\label{table:anomaly-detection-ablation}
	\centering
	\resizebox{0.5\textwidth}{!}{
		\begin{tabular}{llll}
			\toprule
			Method & Precision & Recall & Dice \\
			\midrule
			WeakAnD (thresholding)      	  		& 0.614 (0.05) & 0.504 (0.06) & 0.550 (0.04) \\ 
			WeakAnD (binary)       					& 0.716 (0.07) & 0.620 (0.12) & 0.655 (0.07) \\ 
			WeakAnD (convex-hull) 		        	& 0.708 (0.07) & 0.836 (0.08) & 0.761 (0.04) \\ 
			WeakAnD (w/o closing/opening)       	& 0.727 (0.06) & 0.815 (0.07) & 0.765 (0.03) \\
    		\textbf{WeakAnD}                        & \textbf{0.739 (0.06)} & \textbf{0.808 (0.07)} & \textbf{0.768 (0.03)} \\
			\bottomrule
		\end{tabular}
	}
\end{table}

\begin{figure}[t]
	\centering
	\includegraphics[width=0.4\textwidth]{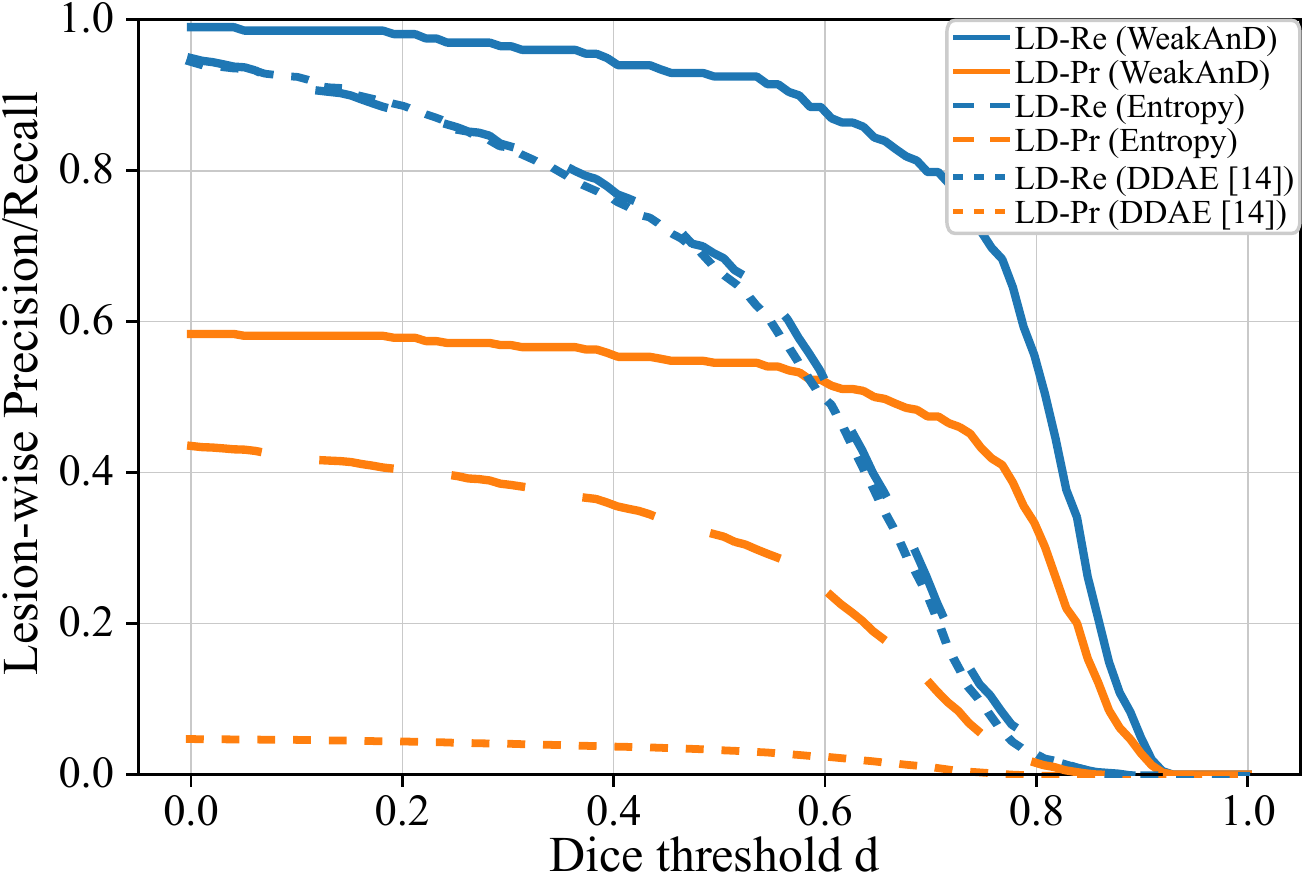}
	\caption{Lesion detection Recall (LD-Re) and Precision (LD-Pr) curves for the proposed approach (solid) and the baseline methods.
	The low LD-Precision curve of \cite{seebock2018identify} can be explained by its noisy segmentation results which lead to several tiny false positive lesions.}
	\label{fig:lesionWiseDetection}
\end{figure}

\paragraph{Lesion-wise Detection}
Lesion-wise precision and recall curves are shown in Fig.~\ref{fig:lesionWiseDetection}. The corresponding curves for the baseline methods are also included for comparison purposes.

\begin{figure*}[t]
	\centering
	\includegraphics[width=0.8\textwidth]{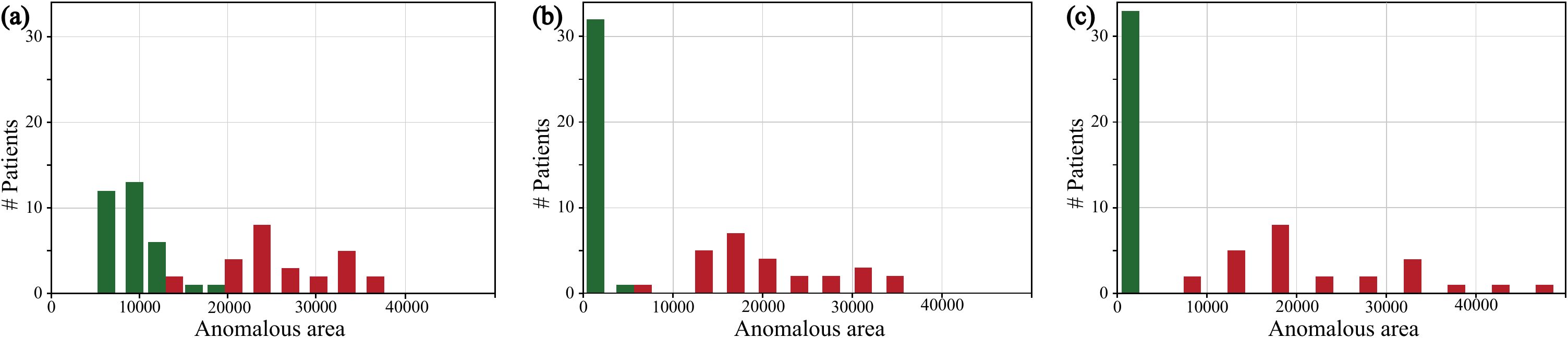}
	\caption{Histograms of the volume-wise classification experiment for detecting late wet AMD cases. Results of (a) DDAE~\cite{seebock2018identify}, (b) entropy of soft predictions and (c) our method. The horizontal axis represents the average number of anomalous pixels per B-Scan for each volume and the vertical axis indicates the number of patients. Green and red denote patients from the \emph{normal evaluation} and the \emph{late wet AMD test} datasets, respectively. Note that a separate overview of the classification results of our method is plotted in Fig.~\ref{fig:healthyVSdiseased}.}
	\label{fig:healthyVSamd}
\end{figure*}

\paragraph{Volume-wise Disease Detection}
Fig.~\ref{fig:healthyVSamd} depict histograms for the volume-wise disease detection experiment, both for the two baselines (\cite{seebock2018identify} (a) and entropy (b)) and our method (c). Red bars correspond to patients from the \textit{late wet AMD} data set, while green bars are associated to patients in the \textit{normal evaluation} set. The horizontal axis represents the average number of anomalous pixels per B-scan for each volume, while the vertical axis indicates the number of patients with a similar anomalous area.
Fig.~\ref{fig:healthyVSamd} (b) and (c) shows no overlap between the healthy and the abnormal sets, while Fig.~\ref{fig:healthyVSamd} (a) does.
Qualitative examples of the anomalies detected in healthy cases from the \textit{normal evaluation} set are depicted in Fig.~\ref{fig:healthyTestSamples}. Both images correspond to the cases with the largest anomalous area. The detected anomalies in these cases correspond to imaging artifacts (Fig.~\ref{fig:healthyTestSamples}, top) or small deviations from normal retinas such small drusen deposits (Fig.~\ref{fig:healthyTestSamples}, bottom). A small false positive is observed at the center of the fovea.

\begin{figure}[t]
	\centering
	\includegraphics[width=0.7\columnwidth]{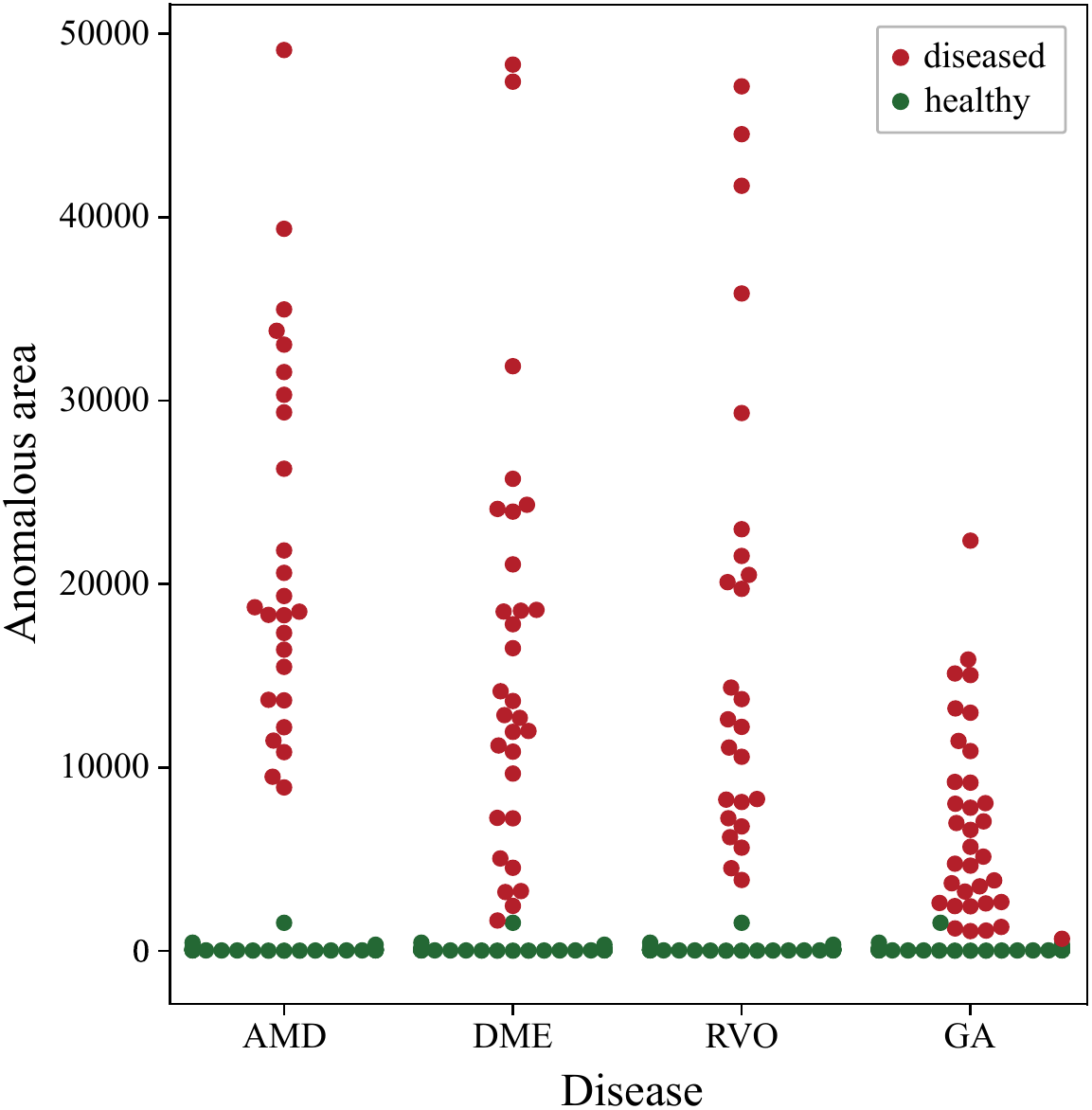}
	\caption{Categorical scatter plot showing the results of all volume-wise classification experiments (AMD, DME, RVO and GA). Each dot represents a patient volume. Diseases are indicated in the horizontal axis, while the vertical axis represents the average number of anomalous pixels per B-Scan for each volume. Green and red denote patients from the \emph{normal evaluation} and the \emph{diseased} datasets, respectively.}
	\label{fig:healthyVSdiseased}
\end{figure}

Fig.~\ref{fig:healthyVSdiseased} presents scatter plots showing the average number of anomalous pixels per B-scan for each diseased/healthy volume in our volume-wise classification experiments. As in Fig.~\ref{fig:healthyVSamd}, it can be seen that this feature is an almost perfect predictor for this application. Qualitative results of the central B-scan of DME, RVO and GA cases, respectively, are presented in Fig.~\ref{fig:healthyVSdiseased_qualitative}. The anomalous region detected in Fig.~\ref{fig:healthyVSdiseased_qualitative} (a) covers parts of the retina with intraretinal cystoid fluid. The segmentation in Fig.~\ref{fig:healthyVSdiseased_qualitative} (b) shows a similar behaviour, although it also includes areas of intraretinal hyperreflective foci. Finally, Fig~\ref{fig:healthyVSdiseased_qualitative} (c) illustrates that our method is also capable of selectively detecting areas of RPE atrophy and neurosensory thinning in eyes with GA. 

\begin{figure}[t]
	\centering
	\includegraphics[width=1.0\columnwidth]{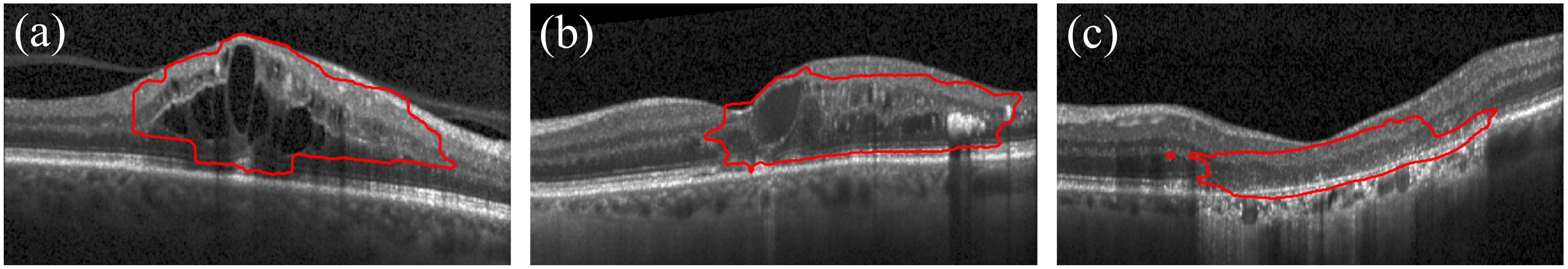}
	\caption{Qualitative results of the proposed method on (a) DME, (b) RVO and (c) GA cases.}
	\label{fig:healthyVSdiseased_qualitative}
\end{figure}

\begin{figure}[t]
	\centering
	\includegraphics[width=0.48\textwidth]{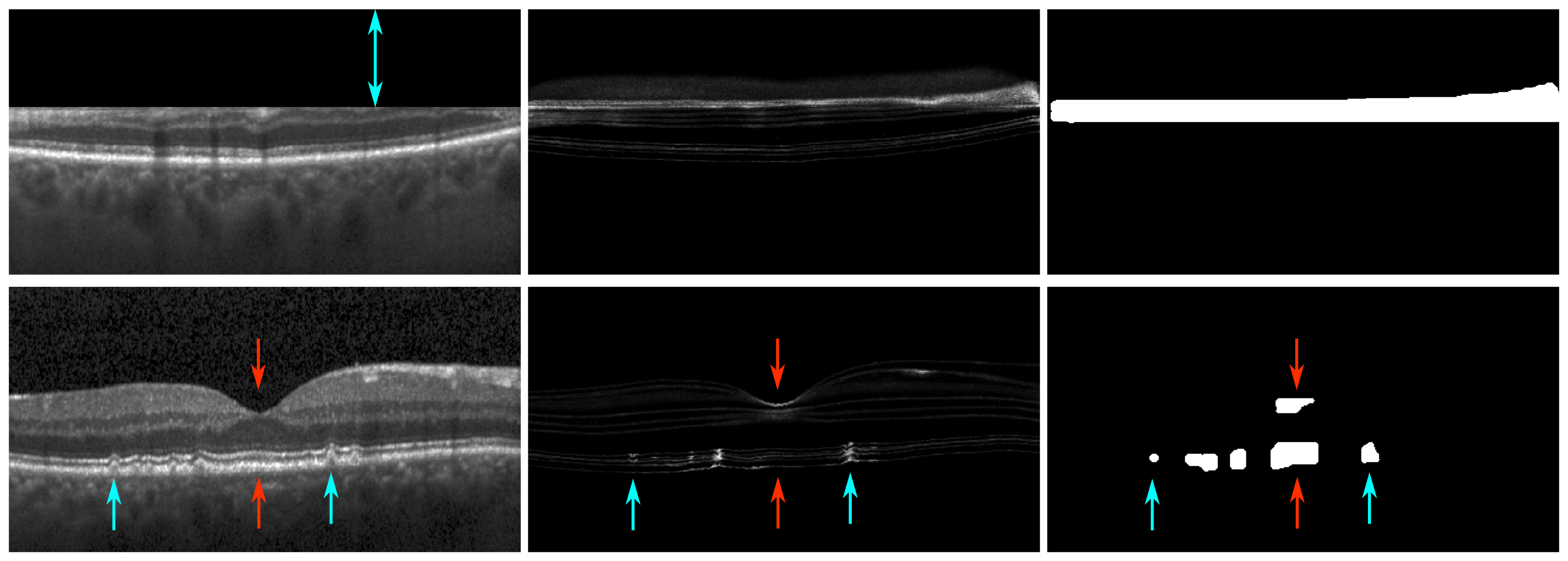}
	\caption{Anomaly detection in normal scans. Two B-scans from the \emph{normal evaluation} dataset with the largest anomalous area are shown. From left to right: Original B-scans, uncertainty maps and corresponding anomaly segmentation results. Top row: cut edge artifact (blue arrow). Bottom row: small drusen (blue arrows) and false positives (red arrows).} 
	\label{fig:healthyTestSamples}
\end{figure}

\section{Discussion}
\label{sec:discussion}
We propose to detect and segment anomalies in retinal OCT images using epistemic uncertainty estimations. The approach is built on the assumption that epistemic uncertainty correlates with unknown anatomical variability (anomalies), not present in the training data. This claim is supported by the results, in particular by the high correlation ($\rho$=0.91) between the amount of anomalous area and uncertainty, as observed in Fig.~\ref{fig:correlation-uncertainty-anomaly}.
Another alternative to identify anomalies is to use the entropy of the soft predictions of the layer segmentation method. Using the soft predictions of neural networks directly has been previously explored as an alternative to identify out-of-distribution samples~\cite{hendrycks2016baseline,ratzlaff2018methods}. We used this idea as a baseline to compare with and we observed that epistemic uncertainties are more powerful to reflect abnormal changes with respect to the training set (Table~\ref{table:anomaly-detection}, Fig.~\ref{fig:lesionWiseDetection}). We believe this is caused by the softmax predictions capturing different information than uncertainty estimates obtained through MC-sampling. The soft predictions indicate the probability of a given pixel belonging to a specific class, while an uncertainty estimate provides information regarding the confidence of the network about assigning a specific likelihood. As pointed out by \cite{gal2015bayesian}, a model can be uncertain in its predictions even with a high softmax output for a specific class. Nevertheless, it is interesting to observe that this baseline is still able to separate between normal and late wet AMD cases based on anomalous area, as observed in Fig.~\ref{fig:healthyVSdiseased}.

We also took advantage of weak supervision by training our segmentation model with labels provided by an existing automated approach~\cite{garvin2009automated}, which is known to perform accurately in healthy scans. Thus, instead of relying on a large training set of normal and diseased patients with costly per-pixel manual annotations of anomalies, we proposed to train our approach using a normal data set, providing anatomical information via weak labels. Our empirical observations showed that using this alternative still results in high performance. Nevertheless, the segmentation U-Net is not limited to be trained with weak labels: we argue that it could also be trained using manual annotations without loss of generality. 

Compared to the baseline method for anomaly detection in OCT~\cite{seebock2018identify}, our approach achieved significantly better results in terms of several quantitative metrics (Table~\ref{table:anomaly-detection}). Interpreting these pixel-wise quality measures requires taking into consideration that manually annotating anomalies is a difficult task: transitions between healthy and diseased scans are continuous, often unclear, hard to define and exposed to subjective interpretation. Therefore, ensuring exact and consistent ground truth labellings is nearly impossible. The high degree of overlap between the outputs of our model and the manual annotations indicates then that the proposed approach is able to approximate the performance of a human expert. This is also supported by the fact that the worst observed Dice value (0.72) is relatively high. 
To complement these interpretations, we also evaluated the performance of the proposed approach to detect lesions as such. The evaluation of the lesion-wise detection experiment, depicted in Fig.~\ref{fig:lesionWiseDetection}, linked quantitative pixel-based evaluation metrics with lesion-level detection capabilities. It can be seen that increasing the requirement of Dice performance $d$ for lesions from $0.0$ to $0.6$ only decreases the lesion detection performance by ~10\%, as measured in terms of lesion detection precision and recall. These results indicate that the ability of the method to accurately identify the borders of the anomalies does not have a significant effect in the lesion detection performance, as most of the overlap area with the human expert annotation is located in affected tissue. In other words, most of the changes in Dice are explained by differences in the borders of the anomalous regions (as seen in Fig.~\ref{fig:qualitative_results}, right column). 

Moreover, it was observed that the size of the predicted anomalous areas was an almost perfect discriminator to classify normal vs. diseased subjects.
We hypothesize that this is a consequence of our method being able to detect abnormalities in diseased subject without oversegmenting false positives in healthy subjects (Fig.~\ref{fig:healthyTestSamples}). Nevertheless, while our method is able to accurately identify abnormal cases from normal based only on the amount of detected anomalous area, further research should be made to evaluate its potential in a screening setting, e.g. by evaluating its discrimination ability in early diseased cases.

Although our method does not rely on ground truth annotations of abnormalities for training, it still has hyperparameters that need to be optimized. In our experiments, we used a late wet AMD validation set comprising 5 OCT volumes. We observed that the method is not too sensitive to changes in the dropout rate (Table~\ref{table:anomaly-detection-validation}). We qualitatively recognized that changes in the amount of dropout mostly affect the magnitude of the uncertainty values, although their distribution in the image remains relatively stable. This effect is mostly absorbed by the post-processing stage, resulting only in slight variations of the final segmentations. On the other hand, our classification results in different diseases indicate that using a validation set with only one specific condition might be enough to ensure good generalization.

Finally, our approach also reported a significantly better performance than the DDAE method~\cite{seebock2018identify} when evaluated both on a lesion and a volume basis. This might be a consequence of the noise in the segmentations generated by the baseline approach, which results in several small isolated regions that increase the false positive rate. Nevertheless, it is important to note that \cite{seebock2018identify} tackles a more difficult task, not only to segment anomalies but to identify categories of them. This restrained such an approach to a localized representation that enables subsequent clustering on local level. In contrast, our method only aims to provide accurate pixel-wise segmentation for the detected abnormalities, not specifically designed for a subsequent clustering step of anomalies. In addition, the method in~\cite{seebock2018identify} does not incorporate anatomical information during training nor post processing.

The qualitative analysis of the results in Fig.~\ref{fig:qualitative_results} revealed that
the uncertainty maps showed high values surrounding subretinal fluid (SRF), a concave form in cases of pigment epithelial detachments (PED) and dense patterns in regions of hyperreflective foci (HRF). In general, the segmentation model predicted the background class with high confidence in large areas of fluid, probably due to missing edges and/or dark appearance in those regions. This observation highlights the necessity of appropriate post-processing to obtain smooth segmentation maps (Table~\ref{table:anomaly-detection-ablation}).

From the ablation study is also possible to conclude that each part of the method is important to ensure accuracy and consistent results. In particular, we observed that using less informative target labels for the segmentation approach e.g., by targeting the whole retina instead of its constitutive layers (\textit{WeakAnD (binary)} in Table~\ref{table:anomaly-detection-ablation}) decreases the performance for anomaly detection (Dice index drop of 14.7\%). We observed that the uncertainty maps produced by the binary alternative were not as detailed and dense as the ones of the proposed method. This caused segmentation shapes inconsistent with the manual anomaly annotations (see Fig.~\ref{fig:ablationStudies-qualitative}(c)), as well as apparent horizontal and vertical gap-artifacts of segmentation areas.
Considering the fact that the cellular components of the retina are arranged in a layer-wise manner~\cite{hildebrand2011anatomy} and pathologies alter their appearance, using retinal layer information proved to be a particularly appropriate way to incorporate anatomical knowledge into the model. This helped to achieve more representative uncertainty values and, therefore, better results. For this particular point, it is important to emphasize the contribution of the post-processing method based on majority-ray-casting. As observed in Table~\ref{table:anomaly-detection-ablation}, replacing this stage by other alternative approaches caused drops in performance. Removing majority-ray-casting and only conducting thresholding of the uncertainty maps (\emph{WeakAnD (thresholding)}) resulted in poor quantitative results, decreasing the Dice index by 28.4\%. This is also reflected in Fig.~\ref{fig:ablationStudies-qualitative}(d), where the exemplary segmentation covers not only the anomalous regions but also some borders between retinal layers. This result was obtained using an optimal threshold ($t=0.03$) selected on the validation set. Although this might compensate for the discontinuous property in the area with true positive anomalies, it brings further layer interfaces to the final segmentation, where a certain degree of uncertainty is also present. Complementing thresholding with a convex-hull based post-processing also caused unwanted artifacts, e.g. in Fig.~\ref{fig:ablationStudies-qualitative}(e), where a small blob in the top right (remaining after thresholding) caused a peculiar segmentation. This is a consequence of the inability of the convex-hull approach to handle multiple non-connected anomalous areas by definition. On the contrary, the anomalous area is better captured when applying our majority-ray-casting method. This indicates the potential of using a relatively straightforward approach combined with an appropriate post-processing step in the context of anomaly detection. This post-processing stage is crucial to achieve a blob-shaped segmentation, not targeting a specific disease appearance. Our approach is intended to help to transfer the layered output of the uncertainty estimates to a compact segmentation surrounding abnormalities, which we believe is the most straightforward way to highlight them in general. This means that majority-ray-casting allows to obtain an easier-to-interpret result. 
The previously published approach is already able to retrieve such a shape (Fig. 3 in \cite{seebock2018identify}), although with significant false positive detections. Our thresholded uncertainty maps, on the other hand, slightly outperform \cite{seebock2018identify} in terms of Dice, but are not able to retrieve such a blob-shape due to the “partial blindness” of the uncertainty estimates. This is line with what can be seen from Table~\ref{table:anomaly-detection} and Table~\ref{table:anomaly-detection-ablation}, where \cite{seebock2018identify} reported lower precision but higher recall than our method. Finally it is worth mentioning that, in addition to fluid related lesions (Fig.~\ref{fig:qualitative_results} and Fig.~\ref{fig:healthyVSdiseased_qualitative} (a)), our approach detects other anomalies such as drusen (Fig.~\ref{fig:healthyTestSamples}, bottom row), hyperreflective material (Fig.~\ref{fig:healthyVSdiseased_qualitative} (b)), DRIL or GA lesions (Fig.~\ref{fig:healthyVSdiseased_qualitative} (c)). This demonstrates that the presented method allows to highlight a variety of retinal abnormalities in multiple diseases.

We observed that the network detected anomalies only in the area ranging from the top of the NFL to the RPE. We believe that this is a consequence of the model being restricted by the anatomy used for training. A similar behavior was observed before in the binary model, trained to segment the retina and the background. By using the weak labels generated using the Garvin \etal~\cite{garvin2009automated} method, our network is unable to capture representative uncertainty estimates in regions that are jointly labeled as background. Our hypothesis is that the network optimizes its loss function by focusing more on the non-background layer labels. This makes the network invariant to changes in areas below the RPE and above the vitreous-macular interface, and therefore does not show uncertainties there. 
Incorporating labels for other layers such as the choroid might allow the model to explicitly learn the normal characteristics of these regions, and thus show higher uncertainty estimates when deviations from this appearance are observed (e.g. due to hypertransmision).

Finally, it is worth pointing out a potential limitation of the majority-ray-casting algorithm, related to the internal distribution and localization of anomalies within the retina. Since the post-processing algorithm assumes that areas surrounded by uncertainties are anomalous (Fig.~\ref{fig:method_overview}), there could be specific clinical scenarios in which this assumption does not hold: e.g. in between three independent anomaly detections (Fig.~\ref{fig:healthyTestSamples}, bottom row, bottom red arrow). Hence, this can lead to oversegmentation. In some cases, we also observed false positives in the fovea depression, caused by a thinning in the top retinal layers (Fig.~\ref{fig:healthyTestSamples}, bottom row, top red arrow). Nevertheless, anomaly detection approaches are needed to reach high levels of sensitivity when applied for screening or detecting pathological areas, and false positives are tolerated to a certain extent. Therefore, oversegmentation might not harm the final application. Moreover, the volume-wise disease detection experiment showed perfect separation between diseased and healthy subjects using only the amount of abnormal area for discrimination.

\section{Conclusion}

We proposed a weakly supervised anomaly detection method based on epistemic uncertainty estimates from a Bayesian multiclass U-Net model, with application in retinal OCT analysis. 
The segmentation approach was trained on a cohort of normal subjects to characterize healthy retinal anatomy. No annotations of the target class (anomalies) were used to learn that model. Instead, we took advantage of the fact that traditional segmentation methods work accurately in well-defined environments such as healthy populations, allowing to easily obtain large amounts of segmented data. Following this perspective, we used an automated method~\cite{garvin2009automated} to generate weak labels for the individual retinal layers. 
During test time, unseen B-scans were processed by the Bayesian network, and Monte Carlo sampling with dropout was used to retrieve epistemic uncertainty estimates. To better exploit its application to segment potential anomalies, a novel post-processing technique based on \emph{majority-ray-casting} was introduced. The results showed the importance of this stage to transfer layered output of the uncertainty estimates to binary masks with a smooth segmentation of retinal abnormalities. Future work should investigate how to integrate this prior in the learning process, reducing the reliance on additional post-processing.

The proposed anomaly detection approach needs only healthy samples for training, detects the deviation from normal by exploiting the injected anatomical information of healthy scans and is therefore--by definition--not limited to a specific disease or pathology. An extensive evaluation using 33 normal and 115 diseased OCT volumes (1617 and 5635 B-scans, respectively) demonstrates that our uncertainty-driven method is able to detect anomalies under several conditions,  outperforming alternative approaches. This makes it a promising tool in the context of biomarker discovery, where the detection and exploration of atypical visual variability is a fundamental task.
In this context, further research is planned to explore the suitability of the presented method in the context of biomarker detection. Furthermore, future work should be focused on evaluating the applicability of our approach in a screening setting.

\ifCLASSOPTIONcaptionsoff
  \newpage
\fi



\bibliographystyle{IEEEtran}
\bibliography{references}

%








\end{document}